\documentclass[pra, 12 pt]{revtex4}
\usepackage[english]{babel}
\usepackage{amsmath}
\usepackage{epsfig}

\begin{document}

\title{\bf DENSITY FUNCTIONAL THEORY AND FREE ENERGY OF INHOMOGENEOUS ELECTRON GAS
}
\author{V.B. Bobrov $^{1}$, S.A. Trigger $^{1,2}$}
\address{$^1$ Joint\, Institute\, for\, High\, Temperatures, Russian\, Academy\,
of\, Sciences, 13/19, Izhorskaia St., 13, Bd. 2. Moscow\, 125412, Russia;\\
emails:\, vic5907@mail.ru,\;satron@mail.ru\\
$^2$ Eindhoven  University of Technology, P.O. Box 513, MB 5600
Eindhoven, The Netherlands}

\begin{abstract}
It is shown that in adiabatic approximation for nuclei the many-component Coulomb system cannot be described on the basis of the grand canonical ensemble. Using the variational Bogolyubov's procedure for the free energy, the Hohenberg-Kohn theorem is proved in the canonical ensemble for inhomogeneous electron gas at finite temperature.
The principal difference between consideration in the framework of quantum statistics in the canonical ensemble and quantum-mechanical consideration of a finite number of particles in infinite volume is established. The problem of universality of the density functional for describing the inhomogeneous electron density in a disordered nuclei field is considered.\\

 PACS number(s): 71.15.Mb, 05.30.Ch, 71.10.Ca, 52.25.Kn \\

\end{abstract}

\maketitle

\section{Introduction}

 In 1964 P.Hohenberg and W.Kohn [1] proved the remarkable theorem, which initiated the new method in statistical physics, i.e, the density functional theory (DFT). The DFT is widely used, first, in the condensed matter theory (see,e.g. [2-6]). As applied to the investigation of properties of the real matter the DFT is based on the consideration of a large number of electrons and nuclei, interacting by the Coulomb law (Coulomb system, - CS) [7].

The large difference in the electron and ion masses in the CS allows the use of the adiabatic approximation (Born-Oppenheimer approximation). As is known, the adiabatic approximation is reduced to the following procedure. At first, the electron wave function has to be found in the field of nuclei which have some fixed positions. The obtained wave functions
and the electron subsystem energy depend parametrically on the coordinates of nuclei. In this case, according to the Hohenberg-Kohn theorem the electron system energy is the functional of the electron density [1]. In the next stage, the electron energy is used as an effective many-particle indirect interaction of the nuclear subsystem.  The electron energy increases with the average distance between nuclei [8]. Therefore, in the electron ground state, electrons stabilize the nuclear subsystem. The internuclear repulsion prevents the whole system from collapse and an equilibrium configuration can exist. The Fermi statistics of electrons plays a crucial role in collapse prevention [9].

Application of the DFT to description of thermodynamic properties of matter was initiated by the work of Mermin [10]. In this paper, the Honenberg-Kohn theorem for the nonhomogeneous electron gas at finite temperature has been proved in the grand canonical ensemble. However, there is a certain problem in constructing the Grand canonical ensemble (in what follows, the grand ensemble) for the CS. In conventional consideration (see, e.g. [11]) the grand ensemble is constructed as a set of canonical ensembles with different numbers of particles of various types in each
canonical term. Furthermore, the summation of these terms over all numbers of particles leads to the grand ensemble. For such a procedure for the Coulomb system containing electrons (index $e$) and nuclei (index $i$) in a volume $V$ at temperature $T$ in each canonical term, the quasineutrality condition should be satisfied.

\begin{eqnarray}
\sum_a e z_a n_a^{(0)}=0, \label{F1}
\end{eqnarray}
where $n_a^{(0)}=<N_a>^{(can)}/V$ is the average particle density, $N_a$ is the operator of the total number of particles of type $a$ with the mass $m_a$, average density $n_a$, and charge $z_a e$,  angle brackets $<...>^{(can)}$ mean averaging over the canonical ensemble. In this case, the numbers of particles are not already independent, and the main advantage of the grand canonical ensemble, which allows calculation of the Green's functions of noninteracting particles, furthermore, construction of the diagram technique perturbation theory in the quantum case, disappears. Therefore, this way for constructing the grand canonical ensemble is absolutely unpromissing for calculation opportunities of the statistical theory.

In the adiabatic approximation, the situation is even worse. In this case, using the grand ensemble, it is in principle impossible to satisfy  the quasineutrality equation (\ref{F1}) for each nucleus configuration, which is the necessary equilibrium condition for the CS [7]. This means that the conventional DFT for inhomogeneous electron gas as a method for describing the properties of matter is possible  only within the canonical ensemble (at least in the adiabatic approximation for nuclei).

In the present paper we generalize finite temperatures in the canonical ensemble the Honenberg-Kohn theorem which was established in [1] for the ground state energy of the inhomogeneous electron gas.

\section{Variational principe for free energy and the Hamiltonian of the CS}

In 1956 N.N.Bogolyubov formulated the variational principle which establishes the upper estimate for the free energy of the system with the Hamiltonian $H=H_0+H_1$
\begin{eqnarray}
F<F_0+<H_1>_0^{can}, \label{F2}
\end{eqnarray}
\begin{eqnarray}
F_0=-T \ln Sp \left\{\exp\left(-\frac{H_0}{T}\right)\right\};\,\; <H_1>_0^{can}=Sp \left\{H_1\,\exp\left(\frac{F_0-H_0}{T}\right)\right\}.\label{F3}
\end{eqnarray}
In this case, the smallness of the Hamiltonian $H_1$ is not supposed (see also [12]). The rigorous proof of relation (\ref{F3}) can be found in [13]. In alternative formulation, the variational principle (\ref{F2}) for the free energy is presented in [14,15] for the Bardeen-Cooper-Schrieffer model in the superconductivity theory (see also [16]). In the second quantization representation, the CS Hamiltonian has the form
\begin{eqnarray}
H^{CS}= H_{ee}+H_{ec}+H_{cc}, \label{F4}
\end{eqnarray}
\begin{eqnarray}
H_{aa}=-\frac{\hbar^2}{2m_a}\int  \Psi^+({\bf r})\nabla^2 \Psi({\bf r})d {\bf r}+\int u_{aa}({\bf r_1}-{\bf r_2})\Psi_a^+({\bf r_1}) \Psi_a^+({\bf r_2}) \Psi_a({\bf r_2})\Psi_a({\bf r_1})d {\bf r_1} d {\bf r_2}, \label{F5}
\end{eqnarray}
\begin{eqnarray}
H_{ec}=\int u_{ec}({\bf r_1}-{\bf r_2})N_e({\bf r_1})N_c({\bf r_2}) d {\bf r_1} d {\bf r_2};\;\;u_{ab}=\frac{z_a z_b e^2}{r}. \label{F6}
\end{eqnarray}
Here $u_{ab}(r)$ is the Coulomb interaction potential for particles of
types $a$ and $b$, $\Psi^+({\bf r})$ and $\Psi({\bf r})$ are the field operators of creation and annihilation, $N_a({\bf r})=\Psi^+({\bf r})\Psi({\bf r})$ is the particle density operator for the type $a$, which in the coordinate representation, is written as
\begin{eqnarray}
N_a({\bf r})=\sum_{i=1}^{N_a} \delta ({\bf r}-{\bf R_a}), \label{F7}
\end{eqnarray}
where ${\bf R_a}$ is the coordinate of the nucleus $a$.
It should be noted that the total number $N_a$ of each particle type in the canonical ensemble is a C-number,
\begin{eqnarray}
N_a= \int N_a({\bf r}) d {\bf r}=<N_a>^{can}. \label{F8}
\end{eqnarray}
The free energy of the quasineutral CS under consideration, containing the fixed number of electrons $N_e$ and nuclei $N_c$ in a volume $V$ is given by
\begin{eqnarray}
F^{CS}= -T \ln Sp \left\{\exp\left(-\frac{H_{CS}}{T}\right)\right\}.\label{F9}
\end{eqnarray}
In the adiabatic approximation the free energy $F^{CS}$ can be written as
\begin{eqnarray}
F^{CS}\simeq -T \ln Sp_c \left\{\exp\left(\frac{F_{ec}-H_{cc}}{T}\right)\right\};\,\; F_{ec}=-T \ln Sp_e \left\{\exp\left(-\frac{H_{ee}+H_{ec}}{T}\right)\right\}.\label{F10}
\end{eqnarray}
The function $F_{ec}$ (\ref{F9}) is the free energy of the electron subsystem in the external field of immobile nuclei. Moreover, the operator of nuclei density $N_c({\bf r})$ in the Hamiltonian $H_{ec}$ (\ref{F10}) is a C-number.

\section{The Hohenberg-Kohn theorem for free energy of inhomogeneous electron gas in the canonical ensemble}

Now let us show that in the canonical ensemble, the inhomogeneous density of interacted electrons (for the definite electron number $N_e$ in a fixed volume $V$) in a certain scalar external field with the potential $\varphi_{ext}({\bf r})$
\begin{eqnarray}
n_e({\bf r})=<N_e({\bf r})>^{(can)} \label{F11}
\end{eqnarray}
singly-valued determines this field (with an accuracy to an insignificant constant).  Let $n_e^{(1)}({\bf r})$ be the inhomogeneous electron density in an external field, which is defined by the scalar potential $\varphi^{(1)}_{ext}({\bf r})$. In this case the free energy $F_e^{(1)}$ of the electron system equals
\begin{eqnarray}
F_e^{(1)}=-T \ln Sp_e \left\{\exp\left(-\frac{H_{ee}+\int\varphi^{(1)}_{ext}({\bf r})N_e({\bf r})d {\bf r}}{T}\right)\right\}.\label{F12}
\end{eqnarray}
In turn, let $n_e^{(2)}({\bf r})$ be the inhomogeneous electron density in an external field, which is defined by the scalar potential $\varphi^{(2)}_{ext}({\bf r})$, and the appropriate free energy $F_e^{(2)}$ is given by
\begin{eqnarray}
F_e^{(2)}=-T \ln Sp_e \left\{\exp\left(-\frac{H_{ee}+\int\varphi^{(2)}_{ext}({\bf r})N_e({\bf r})d {\bf r}}{T}\right)\right\}.\label{F13}
\end{eqnarray}
Let us assume that the same electron density can correspond to different external potentials
\begin{eqnarray}
n_e^{(1)}({\bf r})=n_e^{(2)}({\bf r})=n_e({\bf r});\;\;\; \varphi^{(1)}_{ext}({\bf r})\neq \varphi^{(2)}_{ext}({\bf r})+const. \label{F14}
\end{eqnarray}

According to the Bogolyubov variational principle (\ref{F2}),\,(\ref{F3})
\begin{eqnarray}
F_e^{(1)}<F_e^{(2)}+\int\{\varphi^{(1)}_{ext}({\bf r})-\varphi^{(2)}_{ext}({\bf r})\}n_e^{(2)}({\bf r})d {\bf r},\label{F15}
\end{eqnarray}
\begin{eqnarray}
F_e^{(2)}<F_e^{(1)}+\int\{\varphi^{(2)}_{ext}({\bf r})-\varphi^{(1)}_{ext}({\bf r})\}n_e^{(1)}({\bf r})d {\bf r}.\label{F16}
\end{eqnarray}
Summing up these inequalities (\ref{F15}) and (\ref{F16}) we arrive at the contradiction
\begin{eqnarray}
F_e^{(1)}+F_e^{(2)}<F_e^{(1)}+F_e^{(2)}. \label{F17}
\end{eqnarray}
This means that the initial assumption is wrong. Therefore, one can assert that the inhomogeneous density $n_e({\bf r})$ of interacted electrons (for a fixed full number of electrons $N$ in a certain volume $V$) in the canonical ensemble in a certain scalar field $\varphi_{ext}({\bf r})$ singly-valued defines this potential
\begin{eqnarray}
\varphi_{ext}({\bf r})=\varphi_{ext}({\bf r}, T, V, N_e, \{n_e{(\bf r)}\}). \label{F18}
\end{eqnarray}

Moreover, in the canonical ensemble the free energy (see (\ref{F12}) and (\ref{F13})) and the average density $n_e({\bf r})$ (see (\ref{F11})) of the inhomogeneous electron system  at temperature $T$ is completely defined by the total number of electrons $N_e$, the volume $V$ and the external potential $\varphi_{ext}({\bf r})$
\begin{eqnarray}
F_e =F_e (T, V, N_e, \{\varphi_{ext}({\bf r})\}) ;\;\; n_e({\bf r})= n_e ({\bf r}, T, V, \{{\varphi_{ext}(\bf r)}\}), \label{F19}
\end{eqnarray}
with the normalization condition
\begin{eqnarray}
N_e =\int n_e ({\bf r}) d{\bf r} = <N_e>^{(can)}. \label{F20}
\end{eqnarray}

The above proof shows that the free energy functional $F_e(\{n_e({\bf r})\})$ of the inhomogeneous electron system isthe functional of density  $n_e ({\bf r})$ at fixed temperature $T$ and volume $V$,
\begin{eqnarray}
F_e =F_e (T, V, N_e, \{n_e({\bf r})\}). \label{F21}
\end{eqnarray}
It should be stressed that in general the explicit expression of the free energy functional $F_e(\{n_e({\bf r})\})$ depends on the form of the external potential $\varphi_{ext}({\bf r})$, and on the form of the interparticle interaction $u_{a b}({\bf r})$ (see (\ref{F6})). This means that the "universality" of the density functional is related to the thermodynamic parameters of the system. Therefore, for the fixed external field potential and fixed interparticle interaction, the density functional for the free energy is invariable for arbitrary thermodynamic parameters. The exception is the new phase formation, e.g., liquid-solid transition, etc. The determination of the explicit form of the density functional (\ref{F18}) for the external field $\varphi_{ext}(\bf r)$ has the same level of complexity (see below) as the determination of the density functional (\ref{F21}) for the free energy $F_e(\{n_e({\bf r})\})$.

\section{Electrons in the external field of nuclei: quantum mechanics and quantum statistics}

Let us now pay attention that the inhomogeneous average density profile $n_e({\bf r})$ is caused not only by the existence of the external field $\varphi_{ext}({\bf r})$ but also by the finiteness of the system volume $V$. To eliminate the boundary effects, one should pass to the limit $V\rightarrow\infty$. In this case the definition of the limit
\begin{eqnarray}
\lim_{V \to\infty}\frac {1}{V}\left. \int n_e({\bf r})d{\bf r}\right|_V \label{F22}
\end{eqnarray}
is essential (see (\ref{F20})).

As applied to the problem of electrons in the external field of nuclei the external potential
$\varphi_{ext}({\bf r})$ has the form
\begin{eqnarray}
\varphi_{ext}({\bf r})= \sum_{i=1}^{N_c} {u_{ab}({\bf r}-{\bf R}_i^c)}. \label{F23}
\end{eqnarray}
We take into account that $V$ is very large but finite, and the realpassage to the limit $V \rightarrow \infty$ is possible only after averaging over nuclei coordinates. With this remark, the limit (\ref{F22}),  can be considered in two physically different cases:

Variant ($A$): for $V \rightarrow \infty$, the number of electrons $N_e$ remains finite. This case corresponds to the consideration of atoms, molecules, etc. in the framework of the traditional quantum mechanics in the adiabatic approximation for nuclei (see, e.g., [17]). In this case, the limit (\ref{F22}) equals zero, the conceptions of temperature $T$ and, hence, the free energy $F_e$ become irrelevant. The problem is reduced to the determination of the average energy $E_e^{(0)}$ and average density $n_e({\bf r})$ of the ground state for a given number of electrons in the field of a finite number of the immobile nuclei.

Taking into account the quasineutrality condition
\begin{eqnarray}
\sum_{a =e,c}z_a e N_a=0, \label{F24}
\end{eqnarray}
the expressions for $E_e^{(0)}$ and $n_e({\bf r})$ can be written as
\begin{eqnarray}
E_e^{(0)}= E_e^{(0)} \left(N_e,\,  \{{\bf R}_i^c\}, \,z_c \right) = E_e^{(0)} \left(N_e, \, n_e ({\bf r},\{ {\bf R}_i^c\}), \, z_c \right), \quad  n_e({\bf r}) = n_e\left( {\bf r}, \, N_e, \{ {\bf R}_i^c\}, \,z_c \right). \label{F25}
\end{eqnarray}
Relations (\ref{F25}) respond to the conventional quantum-mechanical problem of the determination of the ground state of electrons localized on one (atom), two (molecule) or more Coulomb centers in infinite space. In the most general formulation the problem comes to the determination of the absolute minimum of the ground state and the corresponding distances between nuclei in the many-center problem. It should be stressed that there is only one length parameter in this problem, i.e., the Bohr radius $a_0=\hbar ^2 /(m_e e^2)$ . Therefore, for the characteristic size  $L_e$ of the density $n_e({\bf r})$ inhomogeneity with the accuracy to the $z_c$ and $N_c$-dependent constant one can write $L_e\sim a_0$.
Taking into account (\ref{F24}), this means $n_e({\bf r})\sim N_c a_0^{-3}$.

Variant ($B$): for $V \rightarrow \infty$, the number of electrons $N_e$ (as well as the number of nuclei $N_c$) proportionally increases in such a way that the average electron density $n_e^{(0)}$ (see (\ref{F1})) remains a finite non-zero value. This consideration responds to the "thermodynamic limit" (see, e.g., [18]) and corresponds to the use of statistical (particularly, canonical) ensembles. Namely, for this description within quantum statistics we use the concept of temperature and free energy. The passage to the thermodynamic limit allows us, instead the so-called "extensive" thermodynamic values, in particular, the free energy $F_e$, to use the "intensive" thermodynamic functions, in particular, the free energy $f_V^e$ of the system under consideration, which is related to the unit volume (see, e.g., [11,18]). Then, taking into account (\ref{F1}),(\ref{F19})-(\ref{F23}), one can write
\begin{eqnarray}
F_e \left( T,V,N_e,\{n_e ({\bf r})\}\right) = \left.
 \int f_V^e \left(T, n_e^{(0)}, \{ n_e({\bf r}, \{ {\bf R}_i^c \}) \} \right)d{\bf r} \right|_V , \label{F26}
\end{eqnarray}

\begin{eqnarray}
n_e \left({\bf r}, \{ {\bf R}_i^c \} \right) =  n_e \left({\bf r}, T, n_e^{(0)}, \{ {\bf R}_i^c \} \right). \label{F27}
\end{eqnarray}

Therefore, the functions $f_V^e $ and $n_e({\bf r}, \{ {\bf R}_i^c \} )$ depend on the thermodynamic parameters, i.e., the temperature $T$ and average electron density $n_e^{(0)}$ in the volume $V$ of the system under consideration. Considering the grand canonical ensemble instead the dependence on the average density $n_e^{(0)}$ we arrive at dependence on the chemical potential of electrons $\mu_e^{(0)}$,
\begin{eqnarray}
\mu_e = \left( \frac {\partial F_e}{\partial N_e}\right)_{T.V}=
\left. \frac{1}{V}\int \left(\frac {\partial f_V^e }{\partial n_e^{(0)} }\right)_T d{\bf r} \right|_V , \label{F28}
\end{eqnarray}
If the external field is absent (in a homogeneous system),  relations (\ref{F26}),(\ref{F27}) takes the form
\begin{eqnarray}
F_e ( T,V,N_e) \equiv  f_V^e \left(T, n_e^{(0)}  \right)V,  \quad  n_e({\bf r}) \equiv n_e^{(0)}. \label{F29}
\end{eqnarray}
For the definite form of the functional $ F_e(T,V,N_e, \{ n_e({\bf r}) \})$, the inhomogeneous density $n_e({\bf r})$ is determined, taking into account (\ref{F18}), from the variational equation for the free energy (see, e.g., [11],[17])
\begin{eqnarray}
\delta F_e =0, \qquad \int \left(n_e({\bf r}) - n_e^{(0)}\right) d{\bf r}=0
\label{F30}
\end{eqnarray}
at fixed values of the variables $T$ and $n_e^{(0)}$.

It is easily seen that, taking into account (\ref{F10}), (\ref{F12}), (\ref{F13}), the variational equation
\begin{eqnarray}
\delta E_e =0, \qquad \int \left(n_e({\bf r}) - n_e^{(0)}\right) d{\bf r}=0,
\label{F31}
\end{eqnarray}
directly follows from Eq.~(\ref{F30}). Here the average energy $E_e$ of the system under consideration in the external field is $\left< H_e \right>^{(can)}$, where $H_e$ is the Hamiltonian of the system in the external field. In turn, the solution of the variational equation  (\ref{F31}) leads to the conclusion that the inhomogeneous density $n_e({\bf r})$ is determined from the relation  (\ref{F11}). This statement is equivalent to the known statement that the variational equation (\ref{F31}) for the energy of the quantum-mechanical system leads to the Schrodinger equation for wave functions and corresponding energy levels [17]. If we now apply the variational equation (\ref{F30}) to the right-hand side of the Bogolyubov inequality (\ref{F2}) we immediately arrive at the formulation of the Hohenberg-Kohn theorem for the free energy, which reads that the external field is the single-valued functional of the inhomogeneous electron density.

Moreover, according to the above consideration, the density functional for the ground state energy $E_e^{(0)}$ (\ref{F25}) of the finite number of electrons in an external field of nuclei in the limit $V\rightarrow\infty$ (quantum mechanics) is principally different from the functional for free energy $F_e$ of the electron system in the external field of nuclei in the thermodynamic limit (quantum statistics), even for the case of strong degeneracy ($T \rightarrow 0$). This principal difference is conditioned by different values of the limit (\ref{F22}).

\section{Localized and delocalized electron states}

From physical reasons [7] one can suppose that in a wide temperatures $T$ range and average electron densities $n_e^{(0)}$ the electron states in the external field of a large but finite number of nuclei can be divided into two groups [19,20,21]:

- the "localized" states (index "loc"), in which electrons and nuclei exist in the form of bounded complexes, as \,"atoms"\,, \,"molecules"\, etc.

- "delocalized"\, states (index "deloc"), in which electrons are spread over the entire system volume.

It should be emphasized that such \,"atoms"\, and \,"molecules"\, are not identical to atoms and molecules conventionally considered in quantum mechanics. The essential difference is conditioned by the opportunity for each of the large (in thermodynamic limit - infinite) number of electrons (due to their identical nature) to occupy the electron state of the fixed \,"atom"\, or \,"molecule"\,. This fact is reflected in the statistical description of the system under consideration. Therefore, the electron states in \,"atoms"\, and \,"molecules"\, depend on the thermodynamic parameters, i.e., the temperature $T$ and average electron density $n_e^{(0)}$ (see,e.g., [22]). Then, according to the definition (\ref{F20}), the average inhomogeneous electron density can be represented in the form
\begin{eqnarray}
n_e ({\bf r}, \{{\bf R}_i^c \}) = n_e^{(loc)} ({\bf r}, \{{\bf R}_i^c \}) + n_e^{(deloc)} ({\bf r}, \{{\bf R}_i^c \}),
\label{F32}
\end{eqnarray}
where $n_e^{(loc)} ({\bf r}, \{{\bf R}_i^c \})$ and $n_e^{(deloc)} ({\bf r}, \{{\bf R}_i^c \})$ are the inhomogeneous densities, responding to the "localized" and "delocalized" states, respectively. Following this division one finds
\begin{eqnarray}
N_e = N_e^{(loc)} + N_e^{(deloc)}, \quad
N_e^{(loc)}= \int n_e^{(loc)} ({\bf r}, \{{\bf R}_i^c \})d{\bf r},
\quad   N_e^{(deloc)}= \int n_e^{(deloc)} ({\bf r}, \{{\bf R}_i^c \})d{\bf r}.
\label{F33}
\end{eqnarray}

Here $N_e^{(loc)}$ and $N_e^{(deloc)}$ are the total numbers of \,"localized"\, and \,"delocalized"\, electrons, respectively. In the thermodynamic limit, relation (\ref{F33}) reads
\begin{eqnarray}
n_e^{(0)}= n_e^{(loc)}+ n_e^{(deloc)}, \label{F34}
\end{eqnarray}
where $n_e^{(loc)} = \left. N_e^{(loc)}\!\right/V$ and $n_e^{(deloc)} =  \left. N_e^{(deloc)}\!\right/V$ are
the average electron densities in the \,"localized"\, and \,"delocalized"\, states, respectively. Their ratio depends on the thermodynamic parameters of the system.

It should be emphasized that the electron states are considered in the irregular field of nuclei [19]. In this case, the role of the eigen quantum numbers is played the energy of the \,"localized"\, (with the discrete spectrum) and \,"delocalized"\, (with the continuous spectrum) states and their spin. In general, the coordinates of nuclei ${R_i^c }$, can be added to the set of quantum numbers for the localized states. The localized states with fixed energies can be localized at different points of the system. Therefore, one can interpret these \,"localized"\, states as degenerate with respect to values $\{{\bf R}_i^c \}$. However, this interpretation has relative character, since the vectors $\{ {\bf R}_i^c \}$ and their components are not eigen functions of any operator, affecting on the electron variables. Strictly speaking, the coordinates ${\bf R}_i^c $ are the parameters appeared due to the use of the standard adiabatic approximation, when nuclei are considering as immobile particles [19]. Taking into account that the electron states with the discrete energy spectrum can be localized on one, two or more nuclei, the electron density $n_e^{(loc)}({\bf r}, \{ {\bf R}_i^c \} )$ can be written as
\begin{eqnarray}
n_e^{(loc)}({\bf r}, \{ {\bf R}_i^c \} )= \sum_i  n_{(1 ) }^{(loc)}({\bf r} - {\bf R}_i^c ) +\sum_{j\neq i}  \sum_{s\neq i,j} n_{(2)}^{(loc)}({\bf r}, {\bf R}_j^c, {\bf R}_s^c ) + \dots,
 \label{F35}
\end{eqnarray}
where $ n_{(1)}^{(loc)}$ and $n_{(2)}^{(loc)}$ are the electron densities corresponding to the electron states, localized on one nucleus ("atom"), two nuclei ("molecule") and so on. As follows from (\ref{F33}), (\ref{F35}) the total number of electrons $N_e^{(loc)}$ in the \,"localized"\, states can be represented as the sum of the total numbers of electrons in one-centrum ("atomic") $N_{(1)}^{(loc)}$, two-centrum («molecular») $N_{(2)}^{(loc)}$, etc. states of the discrete spectrum,
\begin{eqnarray}
N_e^{(loc)} =N_{(1) }^{(loc)}+N_{(2)}^{(loc)}+ \dots , \quad  N_{(\alpha ) }^{(loc)} = \int n_{(\alpha ) }^{(loc)}
d{\bf r}. \label{F36}
\end{eqnarray}
The ratios of the numbers $N_{(1) }^{(loc)}$, $N_{(2) }^{(loc)}$, … depend on the thermodynamic parameters of the system under consideration. In the case, when there are only one-centrum ("atomic") states in the system under consideration, the total number of electrons in the \,"localized"\, states, according to (\ref{F35}),(\ref{F36}) is proportional to the number of nuclei: $N_e^{(loc)}\sim N_c$. In this case, in the thermodynamic limit
$n_e^{(loc)}\sim n_c$.

The \,"delocalized"\, electron states describe the states in the continuous spectrum in the irregular field of a very large (infinite in the thermodynamic limit) number of nuclei. For the self-consistent description of the "localized" (in the one-centrum approximation) and \,"delocalized"\, electron states [20,21], electrons in the \,"delocalized"\, states interact not with \,"bare"\, nuclei, but with \,"ions"\,, which are nuclei dressed by \,"localized"\, electrons.

In this case, to describe the \,"delocalized"\, states, it is natural to apply the perturbation theory to the interaction potential between \,"delocalized"\, electrons and \,"ions"\, [20,21].  Then for the inhomogeneous density $n_e^{(deloc)} ({\bf r}, \{{\bf R}_i^c \})$ of the "delocalized" states, we can write
\begin{eqnarray}
n_e^{(deloc)}({\bf r}, \{ {\bf R}_i^c \} )=   n_e^{(deloc)} + \delta n_e^{(deloc)} ({\bf r}, {\bf R}_i^c),
 \label{F37}
\end{eqnarray}
where $\delta n_e^{(deloc)} ({\bf r}, \{{\bf R}_i^c\})$ is completely defined by the interaction potential between \,"delocalized"\, electrons and \,"ions"\,, and by the response functions of the homogeneous electron liquid, which is characterized by the temperature $T$ and average density $ n_e^{(deloc)}$ in the \,"delocalized"\, states. The examples of the use of the response functions in the density functional theory are given in [1, 4, 23-25].

Thus, we arrive at the conclusion that the problem of establishing of the explicit form of the density functional  (\ref{F18}) in the thermodynamic limit for the external potential of nuclei is the necessary step for finding the free energy functional $F_e$. Even an approximate solution of this problem is very complicated and requires the use of the theory of disordered Coulomb systems.

\section{Conclusions}

According to the above consideration we can assert the following:

(A) The multi-component Coulomb system in the adiabatic approximation should be considered in the canonical ensemble.

(B) The Hohenberg-Kohn theorem is valid for the free energy of inhomogeneous electron gas in an external field for non-zero temperatures.

(C) The universality of the density functional for the free energy is meaningful with respect to the thermodynamic parameters of the system at the fixed external field and interaction potential.

(D) In the DFT, there is a principal difference between the quantum-mechanical consideration of a finite number of particles in the infinite volume ($N_e/V \rightarrow 0$, $V \rightarrow \infty$) and the quantum-statistical consideration in the canonical ensemble with the subsequent passage to the limit $N_e\rightarrow \infty$, $V \rightarrow\infty$ and $n_e^{(0)}=N_e/V \rightarrow constant$.

(E) Establishing the explicit form of the density functional for the external potential of nuclei in the thermodynamic limit is very complicated and requires of the use of the theory of disordered Coulomb systems.
Descriptions of the \,"localized"\, and \,"delocalized"\, electron states are principally different.

\section*{Acknowledgment}

This study was supported by the Netherlands Organization for Scientific Research (NWO), project no. 047.017.2006.007 and the Russian Foundation for Basic Research, project no. 07-02-01464-a.

\end{document}